\documentclass[conference]{IEEEtran}
\IEEEoverridecommandlockouts
\usepackage{cite}
\usepackage{amsmath,amssymb,amsfonts}

\usepackage{graphicx}
\usepackage{textcomp}
\usepackage{xcolor}
\usepackage{algorithm} 
\usepackage{algpseudocode}
\newtheorem{definition}{Definition}
\usepackage{dsfont}
\usepackage{pdfpages}
\usepackage{hyperref}
\newtheorem{problem}{Problem}
\usepackage{lipsum}  
\def\BibTeX{{\rm B\kern-.05em{\sc i\kern-.025em b}\kern-.08em
    T\kern-.1667em\lower.7ex\hbox{E}\kern-.125emX}}

\begin{document}
\title{Optimization of AoII and QAoII in Multi-User Links
\thanks{This work has been supported by TUBITAK Grant 22AG019.}}
\author{\IEEEauthorblockN{Muratcan Ayik\IEEEauthorrefmark{1}\IEEEauthorrefmark{2}, Elif Tugce Ceran\IEEEauthorrefmark{1}, and Elif Uysal\IEEEauthorrefmark{1}\\ \href{https://cng-eee.metu.edu.tr/}{Communication Networks Research Group (CNG)}}
 \IEEEauthorblockA{\IEEEauthorrefmark{1} Electrical and Electronics Eng. Dept., METU,  \IEEEauthorrefmark{2} Aselsan Inc., Ankara, Turkey}}
 
\maketitle
\begin{abstract}
We consider a network with multiple sources and a base station that send time-sensitive information to remote clients. The Age of Incorrect Information (AoII) captures the freshness of the informative pieces of status update packets at the destinations.  We derive the closed-form Whittle Index formulation for a push-based multi-user network over unreliable channels with AoII-dependent cost functions.  We also propose a new semantic performance metric for pull-based systems, named the Age of Incorrect Information at Query (QAoII), that quantifies AoII at particular instants when clients generate queries. Simulation results demonstrate that the proposed Whittle Index-based scheduling policies for both AoII and QAoII-dependent cost functions are superior to benchmark policies, and adopting query-aware scheduling can significantly improve the timeliness for scenarios where a single user or multiple users are scheduled at a time. 
\end{abstract}


\section{Introduction}
In 2022, an estimated 5.3 billion people use the Internet, corresponding to roughly 66\% of the world’s population \cite{b1}. Worldwide developments in intelligent systems and smart devices create an excessive need for valuable and meaningful real-time information updates. This evolving concept of communication systems may require new solutions to describe "the right piece of information" to transfer the valuable parts of the information \cite{b2}. Among the semantic metrics, Age of Information (AoI) \cite{b3} is proposed to measure the freshness of information. The AoI is the amount of time that has passed since the generation of the most recent status update at the destination. Optimizing timeliness of transmission in networks with multiple sources or destinations with respect to AoI has been investigated in a number of previous works \cite{b4, b5, b6, b7, b8, b9}. The study in \cite{b4} derived a low-complexity technique to evaluate AoI on first-come-first-served and last-come first-served systems having multiple users. In \cite{b5}, a threshold-based lazy variant of Slotted ALOHA is proposed when many devices attempt to transmit status updates via a shared medium on a random-access channel, and the time average AoI is calculated. 

AoI is studied in real-life connections in a relatively small number of studies (see, e.g., \cite{b6,b10,b11}). In \cite{b10}, a method for estimating the average AoI without any synchronization was proposed, and the impact of synchronization inaccuracy is discussed on UDP for real networks. The AoI attained over real-life TCP/IP links served by wireless medium access such as WiFi, LTE, etc., and wired Ethernet access was measured in \cite{b11}. In \cite{b6}, AoI and Age of Information at Query (QAoI) metrics are examined to compare the performances of different scheduling methods via software simulations and Software Defined Radio (SDR) testbed. Learning approaches can also be integrated into the AoI metric; in \cite{b7}, a reinforcement learning (RL) approach is proposed to minimize the long-term average AoI for multi-user networks. 

Whittle Index \cite{b12} policies have an essential role in AoI minimization in multi-user systems~\cite{b7,b8,b9}. In \cite{b8}, a policy is identified to minimize the expected weighted sum of AoI in a push-based wireless network with unreliable channels. The closed-form Whittle Index equation is obtained, and the superior efficiency of Max-Weight and Whittle's Index policies as compared to baseline policies is numerically shown. The optimality of a Whittle Index policy in a multiuser setting is analytically shown in \cite{b9}.

A new performance metric is introduced in \cite{b13} named Age of Incorrect Information (AoII), broadening the idea of fresh updates to fresh informative updates. \cite{b14} considers the Mean AoII as a cost for a scenario where the states of the sources cannot be known by the scheduler beforehand, defines a belief value that corresponds to the probability that the information in the monitor is correct and schedule based on the belief value of the states. \cite{b15} studies a system where perfect Channel State Information (CSI) is unavailable, and the issue is to minimize the AoII without proving the indexability by using an indexed priority policy. In pull-based systems, the information collection and utilization rely on a certain query process. In such systems, the age value is only critical at the time the receiver generates a query. \cite{b16} studies a pull-based point-to-point communication model and defines the QAoI, also shows that QAoI aware optimization might dramatically lower the age value for both periodic and stochastic queries.

In this paper, we study AoII as a performance metric for a multi-user status update system in which, at each time, a single user or multiple users are scheduled to transmit the updates over unreliable channels. We derive and propose a closed-form Whittle Index (WI) solution when the receiver knows the value of the state and demonstrate that the WI-based scheduling policy significantly outperforms the baseline policies. In addition, we introduce a novel performance metric to the pull-based status update framework, named the \textit{Age of Incorrect Information at Query (QAoII)}. Similar to QAoI, for QAoII, the information is only relevant for utilization times, and the communication is initiated based on the receiver demand, enabling resource efficiency for massive systems. The closed-form WI is also computed for this pull-based system considering each user's heterogeneous Bernoulli query process. Comparisons of WI-based scheduling policies for AoII and QAoII costs are demonstrated for pull-based and push-based status update systems. 

\section{System Model}

\begin{figure}[h!]
    \centering
    \includegraphics[scale=0.32]{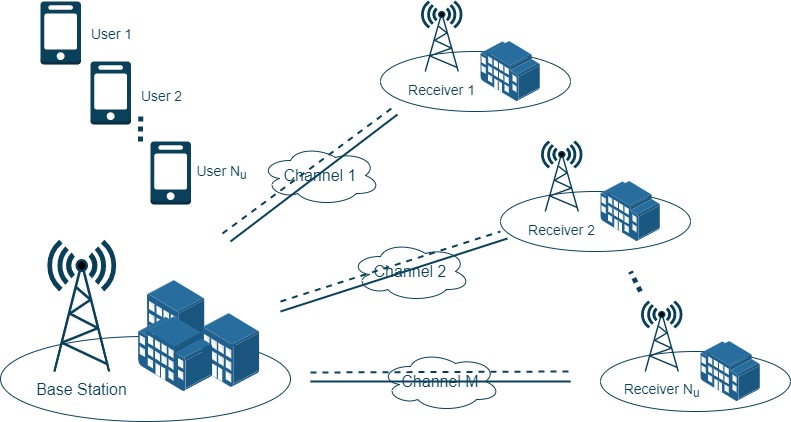}
    \caption{The structure of the system model.}\label{model}
\end{figure}

We consider $N_u$ distinct users that can generate and send their updates to a base station that operates as a central scheduler in discrete time instances. The base station observes the state of the information sources of the users and decides which sources can transmit their updates over unreliable channels such that $M$ ($M \leq N_u$) is the total number of available channels. For each user $i$ in the system, the information process is parameterized by $p_{R_i}$, $N$, and $p_{t_i}$. The source process is assumed to be a Finite State Markov Process with $N$ states. Let $p_{R_i}$ and $p_{t_i}$ denote the probability of remaining in the same state and transitioning to each other particular state for user $i$ for the next frame, respectively. Eq. \eqref{state_relation} relates $p_{R_i}$, $N$, and $p_{t_i}$.
\begin{equation}\label{state_relation}
    p_{R_i}+(N-1)p_{t_i}=1
\end{equation}
 We assume that available channels in the system are independently and identically distributed (i.i.d.) over the frames. The probability of successful transmission for user $i$ is $p_{s_i}$ and $p_{f_i}$ is the probability of unsuccessful transmission for user $i$ which is calculated as $p_{f_i}=1-p_{s_i}$. The structure of the communication model is illustrated in Figure \ref{model}. Dashed lines represent the query case such that the AoII value for a user is valuable only when the query exists for the user. Therefore, the AoII penalty is changed to the QAoII penalty for the cases when the query exists, as explained in Section \ref{query_penalty}. We focus on the case when $p_{R_i}>p_{t_i}$.
\subsection{AoII Optimization}
\label{Penalty Function Dynamics}
In this paper, we investigate two different optimization problems with different penalty functions. Firstly, for the AoII cost \cite{b13}, we aim to minimize the average AoII over an infinite horizon by tracking the AoII value of each user in every frame. Secondly, we add a query state to our system model and aim to minimize the average QAoII of the system and propose low complexity scheduling policies.

We use the linear time-dissatisfaction function and indicator error function, so the penalty of the system is given in \eqref{AoII} as the multiplication of these functions.
\begin{equation} \label{AoII}
    \Delta_{i}(t)=(t-U(t)) \mathds{1}{(\widehat{X_i}(t)\neq X_i(t))} 
\end{equation}
where $U(t)$ is the last time instant where the receiver was in a correct state, ${X_i}(t)$ is a sample of the information process at time $t$ and $\widehat{X_i}(t)$ is an estimation of the process at the receiver for user $i$. The transition probabilities that will be defined are similar to those in \cite[p.~5]{b13}, but for multi-user cases, we have a user index $i$. In the rest of the paper, we use $\Delta_{i}$ and $\Delta_{Q_{i}}$ to denote the AoII and QAoII values for user $i$, respectively. The probabilities can be examined in two cases. 

The penalty function evolution for user $i$ for the case that the receiver has perfect knowledge is the same for both transmission and no transmission cases. The AoII value changes to 0 if the information source does not change its state and 1 if the source changes its state. The probabilities in the transmission case do not depend on the channel because no new information will be transmitted to the receiver due to the perfect knowledge on the receiver side. Unlike the previous case, the penalty function evolution for user $i$ for the case that the receiver does not have perfect knowledge is different for transmission and no transmission cases:

If there is no transmission, the value of the AoII will be equal to 0 if the information source changes its value to the value that is last successfully received by the receiver. The value of the AoII will increase by 1 if the process keeps its same value or transitions to other $N-2$ states. If there is a successful transmission, the value of the AoII will be equal to 0 if the information source does not change; the AoII will increase by 1 if the information source change to other $N-1$ states. However, if the transmission is unsuccessful, the value of the AoII will be equal to 0 if the information source changes its value to the value that is last successfully received by the receiver; the AoII will increase by 1 if the process keeps the same value or transitions to other $N-2$ states.

Let $A_i(t)\in\{0,1\}$ be the binary value corresponding to the decision for user $i$ at time $t$ of the base station to either transmit or remain idle, and $\Delta_{i}(t)$ denote the AoII value for user $i$ at time $t$. The summary of the probability transitions of the AoII penalty function is seen in the following:
\begin{itemize}[
    \setlength{\IEEElabelindent}{\dimexpr-\labelwidth-\labelsep}% Wrapping of text beyond first line of \item
    \setlength{\itemindent}{\dimexpr\labelwidth+\labelsep}% indentation for each new \item
    \setlength{\listparindent}{\parindent}% Restore regular paragraph indentation
  ]
   \item $ \mathds{P}(\Delta_{i}(t+1)=0|\Delta_{i}(t)=0,A_i(t)=0)=p_{R_i}$
   \item $ \mathds{P}(\Delta_{i}(t+1)=1|\Delta_{i}(t)=0,A_i(t)=0)=(N-1)p_{t_i}$
   \item $ \mathds{P}(\Delta_{i}(t+1)=0|\Delta_{i}(t)=0,A_i(t)=1)=p_{R_i}$
   \item $ \mathds{P}(\Delta_{i}(t+1)=1|\Delta_{i}(t)=0,A_i(t)=1)=(N-1)p_{t_i}$
   \item $\mathds{P}(\Delta_{i}(t+1)=0|\Delta_{i}(t)\neq0,A_i(t)=0)=p_{t_i}$
   \item $\mathds{P}(\Delta_{i}(t+1)=\Delta_{i}(t)+1|\Delta_{i}(t)\neq0,A_i(t)=0)=p_{R_i}+(N-2)p_{t_i}$
   \item $ \mathds{P}(\Delta_{i}(t+1)=0|\Delta_{i}(t)\neq0,A_i(t)=1)=p_{R_i}p_{s_i}+p_{f_i}p_{t_i}$
   \item $ \mathds{P}(\Delta_{i}(t+1)=\Delta_{i}(t)+1|\Delta_{i}(t)\neq0,A_i(t)=1)=p_{R_i}p_{f_i}+(N-2)p_{t_i}+p_{s_i}p_{t_i} $
\end{itemize}
Let $\phi$ be the transmission policy defined as a sequence of actions $\phi=(A^\phi(0),A^\phi(1),..)$. Then, the scheduling problem can be formulated as follows:
\begin{problem}[AoII Optimization over Multi-User Links]
\begin{equation} \label{sch}
\begin{aligned}
    \min_\phi  \lim_{T\to\infty} \sup \frac{1}{T} E^{\phi}[\sum_{t=0}^{T-1} \sum_{i=1}^{N_u} \Delta_{i}^{\phi}(t) | \Delta_{i}(0) ] \\ \text{subject to } \sum_{i=1}^{N_u} A_i^{\phi}(t) \leq M 
\end{aligned}
\end{equation}
\end{problem}

\subsection{QAoII Optimization}
\label{query_penalty}
Next, we address the case when information about times at which information gets queried by destination is available at the network server. This scenario allows for optimizing the freshness of the information at the query instants. In this case, the base station will \textit{pull} information from the sources. Thus, as opposed to a push-based model where sources initiate the transmission, it is the base station that initiates it. For this case, we will revise the penalty function to be a combination of query state and AoII penalty. When the query state of a user for a frame is equal to 1, the user is queried; otherwise, the user is not queried. In addition, the QAoII-based cost for user $i$, is calculated as the multiplication of the state of the query and AoII-based cost value for user $i$ in such that:
\[
  \Delta_{Q_{i}}(t) =
  \begin{cases}
    \Delta_{i}(t)     & \text{if $i \in Q(t)$}, \\
    0 & \text{otherwise}.
  \end{cases}
\] 
where $Q(t)$ represents the query set, that is, the set of users with queries at frame $t$. Let $\phi$ be the transmission policy defined as a sequence of actions $\phi=(A^\phi(0),A^\phi(1),..)$. Then, the scheduling problem can be formulated as follows:
\begin{problem}[QAoII Optimization over Multi-User Links]
    \begin{equation} \label{sch_QA}
\begin{aligned}
    \min_{\phi}  \lim_{T\to\infty} \sup \frac{1}{T} E^{\phi}[\sum_{t=0}^{T-1} \sum_{i=1}^{N_u} \Delta_{Q_{i}}^{\phi}(t) | \Delta_{Q_{i}}(0) ] \\ \text{subject to } \sum_{i=1}^{N_u} A_i^{\phi}(t) \leq M 
\end{aligned}
\end{equation}
\end{problem}
\section{Whittle Index Policies to Minimize AoII and QAoII}
Formulation of the AoII-based scheduling problem belongs to the family of Restless Multi-Armed Bandit (RMAB) problems. Reaching the optimal solution to this type of problem is known to be very difficult, so an efficient policy having low complexity called Whittle Index policy can be used \cite{b17, b18}. By relaxing the problem to satisfy the constraint on average, we obtain \eqref{sch_rel}:
\begin{equation} \label{sch_rel}
\begin{aligned}
\min_\phi  \lim_{T\to\infty} \sup \frac{1}{T} E^{\phi}[\sum_{t=0}^{T-1} \sum_{i=1}^{N_u} \Delta_{i}^{\phi}(t) | \Delta_{i}(0) ] \\ \text{subject to}   \lim_{T\to\infty} \sup \frac{1}{T} \sum_{t=0}^{T-1} \sum_{i=1}^{N_u} A_i^{\phi}(t) \leq M
\end{aligned}
\end{equation}
Then, the unconstrained Lagrangian cost \cite{b19} is defined as in \eqref{lagr_mult}:
\begin{equation}\label{lagr_mult}
\begin{aligned}
\lim_{T\to\infty} \sup \frac{1}{T} E^{\phi}(\sum_{t=0}^{T-1} \sum_{i=1}^{N_u} \Delta_{i}^{\phi}(t)+\lambda A_i^\phi(t)|\Delta_{i}(0) )-\lambda M
\end{aligned}
\end{equation}
where $\lambda$ is the Lagrangian parameter and can be considered as the penalty for scheduling users. The Lagrangian problem can be viewed as an infinite horizon average cost Markov decision process (MDP) whose instantaneous cost is defined as $C(\Delta(t), A(t))=\Delta(t)+\lambda A(t)$ for AoII metric, actions represented as $A(t)$, states and transition probabilities were defined previously. As $\lambda M$ in \eqref{lagr_mult} is independent of the scheduling policy $\phi$, it can be eliminated.

$N_u$-dimensional problem can be decomposed into $N_u$ one-dimensional problems that can be solved independently \cite{b20}, so we concentrate on a one-dimensional representation of the problem. For every user in the system, the one-dimensional problem is written as follows:
\begin{equation}\label{lagr_one}
 \text{min}_\phi \lim_{T\to\infty} \sup \frac{1}{T} E^{\phi}(\sum_{t=0}^{T-1} \Delta_{i}^{\phi}(t)+\lambda A_i^\phi(t)|\Delta_{i}(0) )
\end{equation}

\begin{definition}[Threshold policy] \label{thr_def}
A threshold policy is a policy for which there exists a threshold $n$ such that when the current state $\Delta_i < n$, the action is \textit{not transmit} i.e. $A=0$, and when $\Delta_i \geq n$, the action is \textit{transmit} i.e $A=1$, so, $A\in \{0, 1\}$.
\end{definition}

Following the results in~\cite{b13}, a threshold policy parameterized by $n$ exists for a single-user AoII minimization given in~\eqref{lagr_one}. 

\begin{definition}\label{idx_def}
A class is indexable if the cardinality of the set of states in which staying idle \textit{(not transmitting)} is the optimal action increases with the scheduling penalty. When the class is indexable, the Whittle Index can be defined. 
\end{definition}

To establish indexability and to find Whittle Index expressions, the steady-state form of the problem is needed. It can be obtained by resolving the full balance equation under the threshold value \textit{n} at each state as in \cite[Proposition~2]{b13}. The steady-state form of \eqref{lagr_one}, which is the one-dimensional problem under a threshold policy, is:

\begin{equation}\label{ss_thr}
\min_n \quad \quad \overline{\Delta_{i}^n}+\lambda \overline{A_i^n}    
\end{equation}
where $\overline{\Delta_{i}^n}$ is the average value of the cost and  $\overline{A_i^n}$ is the average active time under threshold policy $n$ for user $i$ respectively. 

For a given threshold $n$, the average AoII-based cost formula can be found by $\sum_{k=1}^{\infty}k\pi_{k}(n)$ where $k$ represents the AoII value of a state and $\pi_k(n)$ is the stationary distribution of the Markov chain for a fixed threshold constructed using the transition probabilities given in Section \ref{Penalty Function Dynamics} and is given in \eqref{ave_cost_AoII}.
\begin{equation}\label{ave_cost_AoII}
    \overline{\Delta_i(n)}=\left(N-1\right)p_{t_i}\frac{\frac{1+b_i^n\left(nb_i-n-1\right)}{\left(1-b_i\right)^2}+\frac{b_i^{n-1}a_i\left(n+\frac{1}{1-a_i}\right)}{1-a_i}}{1+\frac{\left(N-1\right)p_{t_i}\left(1-b_i^n\right)}{1-b_i}+\frac{\left(N-1\right)p_{t_i}a_ib_i^{n-1}}{1-a_i}}
\end{equation}
where $a_i$ denotes a constant which is equal to $p_{R_i}p_{f_i}+(N-2)p_{t_i}+p_{s_i}p_{t_i}$ and $b_i$ is a constant which is equal to $p_{R_i}+(N-2)p_{t_i}$. 

The active time corresponds to the portion of time when the transmitter attempts to send packets, so the sum begins at $n$, which is the threshold value and goes to infinity ($\sum_{k=n}^{\infty}\pi_{k}(n)$). Therefore, the average active time formula is given in \eqref{ave_act} for a given threshold.
\begin{equation}\label{ave_act}
    \overline{A(n)}=\frac{\left(N-1\right)p_{t_i}b_i^{n-1}}{\left(1-a_i\right)\left(1+\frac{\left(N-1\right)p_{t_i}\left(1-b_i^n\right)}{1-b_i}+\frac{\left(N-1\right)p_{t_i}a_ib_i^{n-1}}{1-a_i}\right)}
\end{equation}
Defined steady-state probabilities are similar to those in \cite[p.~8]{b13}, but for multi-user cases, we have a user index $i$. Since the steady-state form of the average active time equation, Eq. \eqref{ave_act}, is decreasing with $n$, the one-dimensional problem is indexable (see \cite{b9} for detailed proof) for each user $i$.

After finding the steady state equations and showing the indexability, we can find the closed-form expression of the Whittle Index for each user by computing the ratio of the difference between the average value of the penalty function under threshold values $n+1$ and $n$ and the difference between average active time under threshold values $n$ and $n+1$, where $n$ is equal to the $\Delta_{i}$. Thus, the closed-form Whittle Index equation can be formulated as $\frac{\overline{\Delta_{i}(n+1)}-\overline{\Delta_{i}(n)}}{\overline{A_i(n)}-\overline{A_i(n+1)}}$. Following this, a closed-form WI can be found as in \eqref{whittle_closed}. Note that the detailed derivation of \eqref{whittle_closed} is omitted due to space limitations.
\begin{equation} \label{whittle_closed}
   W_i(\Delta_{i}^n)=\frac{-XY(Z_1-Z_2)}{(a_i-1) (b_i-1)^2 p_{t_i} (b_i-N p_{t_i}+p_{t_i}-1)}
\end{equation} where
\begin{align*}
    X=b_i^{-\Delta_{i}}\Bigr(a_i (N-1) p_{t_i} b_i^{\Delta_{i}}+(a_i-1) b_i^2\\-(a_i-1)b_i ((N-1) p_{t_i}+1)-(N-1) p_{t_i} b_i^{\Delta_{i}+1}\Bigr)\\
    Y=a_i \left((N-1) p_{t_i} b_i^{\Delta_{i}}+b_i-N p_{t_i}+p_{t_i}-1\right)\\-\left((N-1)p_{t_i}b_i^{\Delta_{i}+1}\right)-b_i+N p_{t_i}-p_{t_i}+1\\
    Z_1=\frac{p_{t_i} \left(\frac{a_i (-a_i \Delta_{i}+\Delta_{i}+1) b_i^{\Delta_{i}-1}}{(a_i-1)^2}+\frac{((b_i-1) \Delta_{i}-1) b_i^{\Delta_{i}}+1}{(b_i-1)^2}\right)}{-\frac{a_i (N-1) p_{t_i} b_i^{\Delta_{i}-1}}{a_i-1}-\frac{(N-1) \left(p_{t_i} \left(1-b_i^{\Delta_{i}}\right)\right)}{b_i-1}+1}\\
    Z_2=\frac{p_{t_i}\left(\frac{(b_i \Delta_{i}+b_i-\Delta_{i}-2) b_i^{\Delta_{i}+1}+1}{(b_i-1)^2}-\frac{a_i (a_i \Delta_{i}+a_i-\Delta_{i}-2) b_i^{\Delta_{i}}}{(a_i-1)^2}\right)}{-\frac{a_i (N-1) p_{t_i} b_i^{\Delta_{i}}}{a_i-1}-\frac{(N-1)\left(p_{t_i}\left(1-b_i^{\Delta_{i}+1}\right)\right)}{b_i-1}+1}
\end{align*}

Next, we propose Whittle Index policies for QAoII. We consider the case where query arrivals follow a Bernoulli process that is independent of the age process. The average cost function for QAoII metric can be written as in \eqref{ave_cost_QAoII} leveraging the Bernoulli query arrivals in the summation of Problem \ref{sch_QA}. Therefore, the threshold-based average cost function for QAoII metric is modified such that only a multiplication term is added to the threshold-based average cost function of the AoII metric in \eqref{ave_cost_AoII}. Thus, for the QAoII case, the cost value changes based on the query probabilities of users. Then for a given threshold, the average cost value is formulated as in \eqref{ave_cost_QAoII}.

\begin{equation}\label{ave_cost_QAoII}
    \overline{\Delta_{Q_{i}(n)}}=\sum_{k=1}^{\infty}q_{i}k\pi_{k}(n)
\end{equation}
where $k$ represents the AoII value of a state and $\pi_k(n)$ is the stationary distribution of the Markov chain for a fixed threshold constructed using the transition probabilities given in Section \ref{Penalty Function Dynamics} and the additional query term, $q_i$, is the query probability for user $i$.
Since similar analyses also hold for QAoII metric, we use the cost in \eqref{ave_cost_QAoII} and average active time in \eqref{ave_act} to find the closed form Whittle Index equation, and the function is generated as in \eqref{whittle_closed}.  Note that the threshold value, $n$, is replaced by $\Delta_i$, which corresponds to the AoII value for a user $i$. 

\begin{equation} \label{qwhittle_closed}
   W_i(\Delta_{Q_{i}}^n)=\frac{q_i(a_i-b_i)(X+Y)}{(a_i-1)(b_i-1)^2 (b_i-Np_{t_i}+p_{t_i}-1)}
\end{equation} where 
\begin{align*}
    X = a_i(N-1)p_{t_i}b_i^{\Delta_{i}}+b_i^2\ ((a_i-1){\Delta_{i}}-1)\\-b_i((a_i-1){\Delta_{i}}-1)((N-1)p_{t_i}+2)\\
    Y = (a_i-1){\Delta_{i}}((N-1)p_{t_i}+1)-a_iNp_{t_i}+a_ip_{t_i}\\-(N-1)p_{t_i}b_i^{{\Delta_{i}}+1}-1
\end{align*}

At each frame, for the AoII penalty case, the Whittle Indices of all users in the system are computed by \eqref{whittle_closed} or \eqref{qwhittle_closed} and \textit{M} user(s) having the highest Whittle Index values at this frame are selected to be scheduled as in Algorithm \ref{algo}.
\begin{algorithm}
	\caption{Proposed WI Algorithm for AoII and QAoII}\label{algo}
	\begin{algorithmic}[1]
 \For{t = 0,1,2,\ldots}
 \State Observe the AoII, $\Delta_i(t)$, for all users $i \in \{1,\ldots,N_u\}$.
		\State Compute Whittle Indices for all users by using Eqs. \eqref{whittle_closed} or \eqref{qwhittle_closed} for the optimization of AoII and QAoII, respectively.
        \State Select the \textit{M} users having the highest WI values to schedule.
   \EndFor     
	\end{algorithmic} 
\end{algorithm}

When minimizing the average QAoII of the system, the AoII values at query times for users generating queries are summed, and the summation is divided by the total number of queries as in \eqref{qaoii_equ}.
\begin{equation}\label{qaoii_equ}
    \overline{\Delta_{Q_{i}}}=\frac{\overline{\mathds{1}(i\in Q)\Delta_{i}}}{\overline{\sum_{i=1}\mathds{1}(i\in Q)}}
\end{equation}

\section{Simulation Results}
This section includes the results generated on MATLAB. WI-based scheduling algorithms are tested, and the average AoII and QAoII costs optimized by the WI policy are compared with the baseline Round Robin (RR) and Greedy policy (GP) methods throughout the experiments. We also compare our average AoII results with those generated using the AoI-based WI algorithm in \cite{b8} and for the QAoII case we compare the query-modified version of it. RR algorithm selects a user or multiple users to schedule at each frame in a round-robin fashion. In the GP, user(s) having the highest cost (highest AoII) for the specified frame are selected to schedule. 

\subsection{Results: AoII Optimization}
\begin{figure}[htbp]
\centerline{\includegraphics[scale=.37]{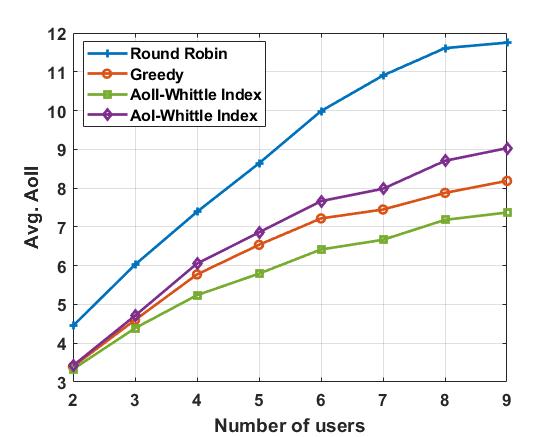}}
\caption{Average AoII values for the Round Robin (RR), Greedy (GP), AoI-Whittle Index (AoI-WI), and AoII-Whittle Index (AoII-WI) scheduling for the system with 2 to 9 users each time a single user is scheduled.}
\label{multi-user-1}
\end{figure}

Figure~\ref{multi-user-1} demonstrates the performance of the proposed WI-based scheduling policy with respect to the number of total users in the experiment allowing a single user to be scheduled each time. The $p_R$ and $p_s$ values are uniformly distributed with respect to the user number of the system, and they are arranged such that while $p_R$ probabilities increase, $p_s$ probabilities decrease. The total frame number is set to 10000. In all cases, the proposed WI based policy is more effective than the GP and RR based scheduling. RR, GP, and WI-dependent average AoII values increase when the number of users increases. Simulation results also show that our proposed WI algorithm is much more efficient in reducing the average AoII on the system compared to using AoI-WI \cite{b8}.

\begin{figure}[htbp]
\centerline{\includegraphics[scale=.37]{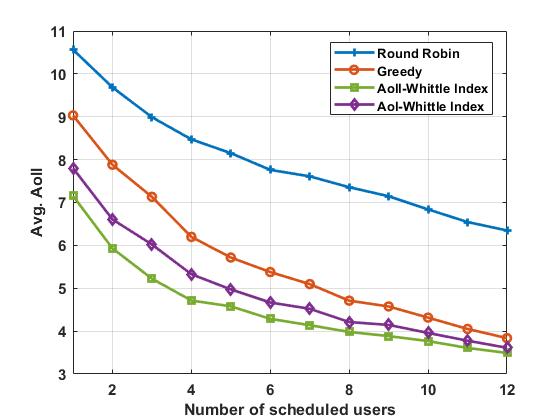}}
\caption{Average AoII values for the Round Robin (RR), Greedy (GP), AoI-Whittle Index (AoI-WI), and AoII-Whittle Index (AoII-WI) scheduling for a system with 37 users, each time the number of scheduled users is increased.}
\label{multi-sch}
\end{figure}

For the experiment results in Figure \ref{multi-sch}, we investigate the average AoII for different numbers of scheduled users at each frame. Without loss of generality, the total number of users is set to 37. The probabilities $p_R$ and $p_s$ for each user are chosen to start at 5\% and increase by 2.5\% up to 95\%, and the total frame number is set to 2000.  The proposed WI policy is more effective than the GP and RR. Our proposed WI  algorithm is also efficient in reducing the average AoII on the system compared to the average AoII generated using AoI-WI. Note that RR, GP, and WI-dependent average AoII values decrease when more users can be scheduled at each frame. Also, when the number of scheduled users increases, the performance gap between different scheduling policies reduces. 
\subsection{Results: QAoII Optimization}
\begin{figure}[htbp]
\centerline{\includegraphics[scale=.37]{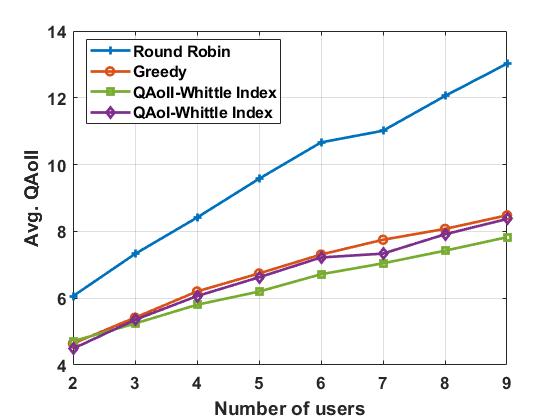}}
\caption{Average QAoII values for the Round Robin (RR), Greedy (GP), QAoI-Whittle Index (QAoI-WI), and QAoII-Whittle Index (QAoII-WI) scheduling for the system with 2 to 9 users each time a single user is scheduled.}
\label{q-multi-user-1}
\end{figure}

Figure~\ref{q-multi-user-1} illustrates the performance of the proposed WI-based scheduling policy with respect to the number of total users in the experiment allowing a single user to be scheduled each time. The $p_R$, $p_s$, and the query probabilities are uniformly distributed with respect to the user number of the system, and they are arranged such that while $p_R$ and query probabilities increase, $p_s$ probabilities decrease. In the simulation, the total frame number is set to 1000, and 25 Monte Carlo simulations are conducted.  In all cases, the proposed WI based policy is more effective than the GP and RR based scheduling. Also, our proposed WI algorithm is much more efficient in reducing the average QAoII on the system compared to the query-modified version of AoI-WI. Note that RR, GP, and WI-dependent average QAoII values increase with increasing the number of users in the system. 

\begin{figure}[htbp]
\centerline{\includegraphics[scale=.37]{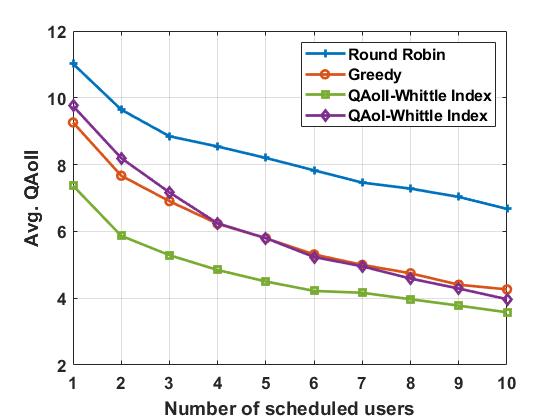}}
    \caption{Average QAoII values for the Round Robin (RR), Greedy (GP), QAoI-Whittle Index (QAoI-WI), and QAoII-Whittle Index (QAoII-WI) scheduling for a system with 37 users, each time the number of scheduled users is increased.}\label{q-multi-sch}
\end{figure}

Figure \ref{q-multi-sch} shows the comparison of the performance of the proposed WI based policy to baseline RR and GP scheduling policies with respect to the number of scheduled users for a 37 users system. The query state is considered throughout the experiment. The selected user to be scheduled is increased by 1 for each run. $p_R$ and $p_s$ probabilities are arranged such that they start at 5\% and increased by 2.5\% up to 95\%. The query probabilities for the users are set in that they start at 95\% and decreased by 2.5\% to 5\%. The total number of frames in this experiment set is selected as 2000. Figure \ref{q-multi-sch} shows the simulation results of this experiment. The proposed query-based WI policy is more effective in all cases than the GP and RR scheduling. RR, GP, and WI-dependent average QAoII values decrease when the scheduled user number increases. Our proposed Whittle Index algorithm also efficiently reduces the average QAoII on the system compared to the average QAoII generated by the query-modified version of AoI-WI. Also, when the number of scheduled users increases, the difference between GP and WI policies decreases. 

For the experiment set results in Table \ref{tab:aoii_set2}, there are 3 users in the system having the parameters for User 1: $p_R=5\%$, $p_s=95\%$, for User 2: $p_R=50\%, p_s=50\%$ and for User 3: $p_R=95\%, p_s=5\%$. The central scheduler selects a single user at each frame to be scheduled based on RR, GP, and WI scheduling algorithms considering AoII and QAoII costs throughout a total of 2000 frames. The average AoII values for this set are presented at the top of Table \ref{tab:aoii_set2}. Also, for the same $p_R$ and $p_s$ probability of users, a query probability is added to the system such that it is 20\%, 50\%, and 80\% for users 1, 2, and 3, respectively. The average QAoII values for this set are summarized at the bottom of Table \ref{tab:aoii_set2}. Numerical results show that
query-aware scheduling can significantly reduce the average AoII experienced by the receiver and
 higher timeliness can be achieved for pull-based systems. The decrease in age values computed by AoII and QAoII metrics is approximately 32\% for GP and approximately 14\% for WI-based scheduling.
\begin{table}[htbp]
\caption{Average AoII \& QAoII values for different scheduling policies for a system with 3 users when a single user is scheduled.}
\begin{center}
\begin{tabular}{|c|c|c|c|}
\hline
\textbf{Avg. age} & \textbf{RR} & \textbf{GP}& \textbf{WI} \\
\hline
 AoII & 12.587 & 11.726 &  7.820  \\
\hline
 QAoII & 9.035  &  7.924 & 6.765  \\
\hline
\end{tabular}
\label{tab:aoii_set2}
\end{center}
\end{table}
\section{Conclusion}
We considered a multi-user uplink system with unreliable channels. We propose closed-form Whittle Index policies for AoII and QAoII cost functions and compared the performance with benchmark policies. Simulation results show that the Whittle Index based scheduling policy is superior in various settings. In future studies, theoretical analyses and performance guarantees of the proposed algorithms will be investigated.

\end{document}